# A strategy for the design of skyrmion racetrack memories

R. Tomasello,[1] E. Martinez,[2] R. Zivieri,[3] L. Torres,[2] M. Carpentieri,[4] and G. Finocchio[5,*]

[1]Department of Computer Science, Modelling, Electronics and System Science, University of Calabria, Via P. Bucci I-87036, Rende (CS), Italy.

[2]Department of Fisica Aplicada, Universidad de Salamanca, Plaza de los Caidos s/n, E-38008, Salamanca, Spain.

[3]Department of Physics and Earth Sciences and CNISM Unit of Ferrara, University of Ferrara, Ferrara, via Saragat 1, I-44122 Ferrara, Italy

[4]Department of Electrical and Information Engineering, Politecnico of Bari, via E. Orabona 4, I-70125 Bari, Italy.

[5]Department of Electronic Engineering, Industrial Chemistry and Engineering, University of Messina, C.da di Dio, I-98166, Messina, Italy.

*Correspondence to gfinocchio@unime.it

Magnetic storage based on racetrack memory is very promising for the design of ultra-dense, low-cost and low-power storage technology. Information can be coded in a magnetic region between two domain walls or, as predicted recently, in topological magnetic objects known as skyrmions. Here, we show the technological advantages and limitations of using Bloch and Néel skyrmions manipulated by spin current generated within the ferromagnet or via the spin-Hall effect arising from a non-magnetic heavy metal underlayer. We found that the Néel skyrmion moved by the spin-Hall effect is a very promising strategy for technological implementation of the next generation of skyrmion racetrack memories (zero field, high thermal stability, and ultra-dense storage). We employed micromagnetics reinforced with an analytical formulation of skyrmion dynamics that we developed from the Thiele equation. We identified that the excitation, at high currents, of a breathing mode of the skyrmion limits the maximal velocity of the memory.

The continual miniaturization of CMOS (complementary metal-oxide semiconductor) devices has led to different challenging aspects such as increased power dissipation and device stability. Moore's law for CMOS is approaching fundamental physical limits imposed by heating and the Heisenberg uncertainty principle. With this in mind, there is an increasing effort to identify alternative or hybrid technologies as replacement for CMOS. At the current juncture, non-volatile memories and logics based on nanomagnetic elements have shown great promise in this regard. Nanomagnetic logic have the potential to bring together real time magnetic reconfiguration and universal memory with high thermal stability and non-volatile nature[1]. A promising magnetic memory architecture is the racetrack initially proposed by IBM[2], where the information bit can be coded in magnetic regions separated by domain walls (DWs), which can be moved by means of the spin-transfer-torque (STT). Despite the initial promises of the DW racetrack memory, its technological implementation has been limited by pinning resulting from the presence of defects (edge and/or surface roughness, dislocations, imperfection etc.)[3], the Walker limit[4] and the Joule heating. Those limits have been partially overcome by the spin-Hall effect[5,6,7] (SHE) and the Dzyaloshinskii–Moriya interaction[8,9] (DMI) in high perpendicular materials[10,11]. Particularly, the interfacial DMI stabilizes Néel walls with a given chirality different to the Bloch wall configuration expected from magnetostatics, while the spin current generated by the SHE gives rise to a Néel DW motion at higher velocities (∼400m/s) and lower current with respect to the Bloch DW one (it can be moved only by the STT)[11].

It has been also predicted that the DMI should be the key ingredient for the nucleation of skyrmions[12, 13, 14,15, 16, 17, 18, 19]. These topologically protected magnetic configurations can code information, and can be the basis of a new type of racetrack memories[14, 21]. They should be ultra-dense and low power consumption, however a detailed analysis of the technological potential is nowadays missing and very demanding.

A skyrmion racetrack memory can be obtained in four different scenarios (A), (B), (C) and (D) by combining the skyrmion type, Bloch[13] (azimuthal magnetization in the skyrmion boundary) or Néel[20] (radial magnetization in the skyrmion boundary), and its motion driven by the STT or the SHE. Fig. 1 summarizes the four different scenarios and the insets show the skyrmion types. In scenarios A, C, and D the magnetic strip is 1000x100x1nm$^3$, while in B it is 100x1000x1nm$^3$. A Cartesian coordinate system has been also introduced. The current flows always along the positive *x*-direction.

Skyrmions (Néel/Bloch) are stabilized by the DMI (interfacial/bulk). The interfacial DMI, observed in bi-layer ferromagnet/heavy metal, is linked to the inversion-symmetry breaking at the interface in the presence of strong spin-orbit coupling[21, 22, 23] (scenarios A and B). The bulk or volume DMI has its origins in the breaking of inversion symmetry within a unit cell of a non-centrosymmetric crystal with non-negligible spin-orbit interactions[24, 25, 26] (scenarios C and D). The STT exerts torques (adiabatic and non-adiabatic) on skyrmions proportional to the current flowing into the ferromagnet ($\mathbf{j_{FE}}= [j_{FE}, 0, 0]$) and to the local magnetization gradient[27] (scenarios A and C). The SHE is linked to the charge current ($\mathbf{j_{HM}} = [j_{HM}, 0, 0]$) flowing through the heavy metal (*x*-direction), which is converted in a spin-current polarized along the *y*-direction because of spin-orbit scattering (scenarios B and D)[28]. The torque exerted on the magnetization by spin currents generated by the SHE is modeled in the Slonczewski form[6, 29, 30, 31]. In all the scenarios, the skyrmion velocity has both a parallel (longitudinal) and a perpendicular (transversal) component to the electrical current flow[17, 32]. In Fig. 1 we indicate the direction of the drift velocity (the larger velocity component). Details about physical and geometrical parameters can be found in Methods. The Bloch skyrmion has been nucleated in agreement with the MnSi phase diagram[16]. The Néel skyrmion has been nucleated by using the same method as described in[21].

## Results

**Comparison of skyrmion motion in the four different scenarios.** Fig. 2a shows the drift velocity of the skyrmion, computed by means of micromagnetic simulations (see Methods), as a function of the current $j_{FE}$ (scenarios A, C) and $j_{HM}$ (scenarios B, D). Those results are achieved for an out-of-plane external field $H_{ext}$ of 250mT. The external field is necessary to nucleate Bloch skyrmions for the scenarios C and D studied here. The skyrmion velocity is proportional to the injected current with different slopes depending on the scenario. In particular, the Bloch skyrmion velocity driven by the SHE (scenario D) exhibits the largest tunability on current. The STT moves both types of skyrmions (scenarios A and C) mainly along the direction of the electrical current (*x*-axis). The origin of this motion is related to the torque terms $\left(\mathbf{j_{FE}} \cdot \nabla\right)\mathbf{m}$ (**m** is the normalized magnetization), which are non-zero along the current direction only. On the other hand, the SHE moves Néel and Bloch skyrmions with the drift velocity in different directions, namely the former (scenario B) along the *y*-axis and the latter (scenario D) along the *x*-axis. This is the first important result of our study. The physical origin of the different directions for the drift velocity for Néel and Bloch skyrmions can be heuristically understood by considering the SHE as source of anti-damping[6, 7, 31]. For a current flow along the $+\hat{x}$-axis, the electrons flow is along the $-\hat{x}$-axis, the spin polarization is along the $+\hat{y}$-axis (the spin-Hall angle is negative) and the SHE stabilizes the magnetization component parallel to the spin-current polarization (the spin current diffuses from the heavy metal to the ferromagnet). Fig. 2b represents a sketch of the mechanism which provides the motion of the Néel skyrmion along the *y*-direction. The SHE reverses the in-plane component of the magnetization (arrow in the square) from -*y* to +*y* and, assuming rigidity, the magnetization reversing is achieved via a skyrmion shift (see supplementary materials MOVIE 1). The same mechanism occurs for a Bloch skyrmion, but in this case because of the azimuthal chirality, the shift is along the *x*-direction (see Fig. 2c) (see supplementary materials MOVIE 2).

The Thiele equation[33] for the scenario B has been derived for a comparison with the numerical results. The "hedgehog" like topological magnetic texture of a Néel skyrmion can be parameterized as $\mathbf{m}(x,y) = \sin\theta(\rho)\cos\phi_0 \hat{\rho} + \sin\theta(\rho)\sin\phi_0 \hat{\phi} + \cos\theta(\rho)\hat{z}$ by setting $\phi_0 = 0$ [34]. The translational motion is obtained by projecting the Landau-Lifshitz-Gilbert (LLG) equation including the SHE onto the relevant translational modes, yielding[33]:

$$\mathbf{G} \times \mathbf{v} - \alpha_G \vec{\mathcal{D}} \cdot \mathbf{v} + 4\pi B \vec{\mathcal{R}}\left(\phi_0 = 0\right) \cdot \mathbf{j_{HM}} = 0 \qquad (1)$$

Here, the vector **G** can be identified as the "gyrocoupling vector", $\mathbf{v}=[v_x,v_y]$ is the velocity of the skyrmion, the matrix $\ddot{\mathcal{D}}=\begin{pmatrix} \mathcal{D} & 0 \\ 0 & \mathcal{D} \end{pmatrix}$ is the dissipative tensor describing the effect of the dissipative forces on the moving magnetic skyrmion, $\ddot{\mathcal{R}}$ is the in-plane rotation matrix. The coefficient $\alpha_G$ is the Gilbert damping and the coefficient $B$ is linked to the SHE (see Supplementary Materials S1). From equation (1), a simple expression of the skyrmion velocities can be derived:

$$\begin{cases} v_x = \dfrac{\alpha_G \mathcal{D} B}{1+\alpha_G^2 \mathcal{D}^2} j_{HM} \\ v_y = \dfrac{B}{1+\alpha_G^2 \mathcal{D}^2} j_{HM} \end{cases} \quad (2)$$

Both components of the velocity are proportional to $j_{HM}$. However, being $v_x$ also proportional to the Gilbert damping ($\alpha_G \ll 1$), it follows that $v_x \ll v_y$. This result confirms the micromagnetic predictions. On the other hand, the Thiele equation derived for the scenario D related to the Bloch skyrmion $(\phi_0 = \pi/2)$ shows that $v_y \ll v_x{}^{32}$, confirming again our numerical results. A complete description of the derivation of the Thiele and the velocity equations for Néel and Bloch skyrmion driven by the SHE can be found in Supplementary Material S1.

The results of Fig. 2a also indicate that for the scenarios A, B and C, the skyrmion is stable for a wide range of current <120 A/cm². Considering the four scenarios, although the scenario D seems to be the most promising configuration from a technological point of view (larger slope of the velocity vs. current curve), it is stable up to $j_{HM}$=15 MA/cm² and at low temperature[35, 36]. Scenario C has been studied both theoretically and experimentally[16, 24, 35]. On the other hand, the experimental implementations of the scenarios A, B and D have not been achieved yet, representing challenges and routes in the design of magnetic materials and devices.

The analysis of the four scenarios leads to the conclusion that one of the most promising strategies for the design of a skyrmion racetrack memory is the framework of the scenario B (Néel skyrmion, SHE) when considering both the skyrmion velocity vs. current scaling and the relative ease of creating these systems (i.e. these systems can be created with standard magnetron-sputtering or evaporation deposition techniques and can be polycrystalline)[11, 37, 38]. For this reason, in the rest of this paper we will focus on the scenario B. Fig. 2d shows the drift velocity as a function of the current in presence of thermal fluctuations $T$=350 K and for both a perfect and rough strip. We consider an edge roughness by randomly removing regions from the strip boundaries with an uniform probability distribution characterized by the typical roughness size $D_g$=4 nm[39]. Those results show that the skyrmion motion is qualitatively not affected by thermal fluctuations and roughness. Quantitatively, the maximal applicable current reduces to 100 MA/cm$^2$ and, in presence of the roughness, the threshold current to move the skyrmion increases to 10 MA/cm$^2$.

**Néel skyrmion motion driven by the SHE at zero field.** We have also performed a systematic study as a function of the external out-of-plane field $H_{ext}$ obtaining no qualitative differences in terms of skyrmion velocity vs. current. Fig. 3a shows the stability phase diagram of the Néel skyrmion as a function of the DMI parameter and the $H_{ext}$. The coloured part of the diagram identifies the region where the skyrmion is stable. It is interesting to note how, reducing $H_{ext}$, the range of the DMI where the skyrmion is stable is also reduced. The line inside this region indicates the values of the DMI and $H_{ext}$ which maintain the skyrmion diameter constant at 46 nm. In order to reduce the power dissipation and to make the design of the skyrmion racetrack memory much simpler, one of the most interesting solutions should be achieved with no bias field (the DMI parameter is 1.8 mJ/m$^2$). For a fixed thickness, the DMI parameter can be controlled experimentally by changing the thickness or the composition of the heavy metal[40, 41] and the thickness of the ferromagnet[21].

Fig. 3b shows the skyrmion velocity as a function of $j_{HM}$ at zero field, zero temperature and perfect strip (black curve). It increases linearly with the current up to 117 m/s at $j_{HM}$=60 MA/cm$^2$. Fig. 3b also shows the skyrmion velocity as computed from the equation (2) (green line with triangles). The quantitative accord between the analytical and the micromagnetic results is excellent, indicating a simple tool for a preliminary design of skyrmion racetrack memories (scenarios B and D). The comparison between the skyrmion velocity as a function of the current for different field configurations (compare Fig. 2d with Fig. 3b) points out that the slope does not change (skyrmion size is constant), while reductions of the maximal applicable current are observed up to 60 MA/cm$^2$ at zero external field. The reduction of the current working region is linked to the transient breathing mode[42] of the skyrmion excited by the application of the current. In particular, for a fixed current, the expansion of the skyrmion increases as $H_{ext}$ decreases (the out-of-plane field stabilizes the magnetization state out of the skyrmion) (compare MOVIE 3 and MOVIE 4). Fig. 3c illustrates the time domain evolution of the skyrmion diameter for three different currents (30, 40 and 50 MA/cm$^2$). The maximal expansion of the skyrmion increases with the current, reaching the boundary of the strip at $j_{HM}$=65 MA/cm$^2$; as consequence, the skyrmion state disappears giving rise to a complex magnetic pattern (see MOVIE 5). This is the second important result of our study, underlining the technological limit of a skyrmion racetrack memory. In other words, at high currents, the information stored in the skyrmion can be lost. Fig. 3d also shows a comparison of the skyrmion velocity as a function of the current for an ideal strip ($T$=0 K) (black curve) and a rough strip and room temperature ($T$=350 K) (red curve). Again, the thermal fluctuations and the roughness do not qualitatively affect the skyrmion motion.

**Néel skyrmion motion driven by the SHE at zero field: effect of the confinement.** We now wish to highlight the differences in our results of scenario B with a previous work[21] detailing a scheme for skyrmion motion. In that work, the motion of the Néel skyrmion driven by the SHE is in the same direction of the applied electrical current and it is achieved for current densities $\leq$ 5 MA/cm$^2$. This seems to contradict both our analytical theory and the results from micromagnetics in scenario B where the skyrmion motion is orthogonal to the current flow direction. With this in mind, we studied the Néel skyrmion motion driven by the SHE, but now in a strip 1000x100x1 nm$^3$ (see Fig. 4a scenario B*). Fig. 4b shows the $x$-component of the skyrmion velocity as a function of the $j_{HM}$. At zero temperature and for a perfect strip (red curve), the threshold current density is 0.01 MA/cm$^2$, while the skyrmion motion is achieved up to 0.5 MA/cm$^2$ ($v_x$=22.8 m/s). This is in accordance with the previous result. However, one can also see that the $y$-position changes during the course of the trajectory and moves closer to the edge of the wire. Our results show that the $y$-position $\Delta y_2$, that the skyrmion equilibrates at during the course of its $x$-motion, is dependent on the current. At larger current the skyrmion is expelled from the strip (see MOVIE 6). The current range over which the skyrmion is stable and the current at which the skyrmion is expelled from the wire seems to depend heavily on thermal fluctuations ($T$=350K) and wire edge roughness (Fig. 4b black curve). Edge roughness tends to slow down the skyrmion and to facilitate its expulsion. These facts suggest that the $x$-directional motion of the skyrmion and the $y$-position of the skyrmion in the wire arises from an interplay among the SHE, magnetostatic confinement[34] and boundary conditions imposed by the DMI[22]. The role of the boundaries in skyrmion motion constitutes the key difference between the two skyrmion motions in B and B*. Equation (2) for the velocity derived for the scenario B could, in fact, be generalized to the scenario B* by taking into account the confinement effect (additional term in equation (1))[34].

We have further analyzed the effect of the confinement on the skyrmion motion in B* for different widths $w$ of the strip. Fig. 4c shows the $x$-component of the skyrmion velocity as a function of $j_{HM}$ for three values of $w$ (100, 150 and 200 nm). The three curves overlap indicating that the confinement effect gives rise to a steady-state velocity independent on $w$. When the current is injected, the skyrmion moves from the equilibrium position (at a distance $\Delta y_1$ (half width), see Fig. 4a), along $y$ and $x$. After a transient time $\Delta t_1$, $v_y$=0 m/s and the skyrmion stabilizes at a distance $\Delta y_2$ from the boundary. $\Delta y_2$ decreases as the current increases. The transient time is around 30 ns for the current range studied, whereas the travelling distance $\Delta x_1$ during this transient time increases linearly with the current, from 74nm at $j_{HM}$=0.01MA/cm$^2$ to 406nm at $j_{HM}$=0.5MA/cm$^2$. In addition, the distance $\Delta y_2$ is independent on $w$ for $j_{HM} > 0.1$ MA/cm$^2$. The existence of the transient $\Delta t_1$ introduces an additional bottleneck in this kind of racetrack memory.

## Discussion

We have micromagnetically analyzed four scenarios (A, B, C and D) for the design of a skyrmion racetrack memory by combining skyrmion type (Bloch or Néel) and driving force (STT or SHE). The fundamental aspect is that we have considered "state of the art" material parameters (see Table 1 in Methods for the parameters details) in order to stabilize Bloch (bulk DMI) or Néel (interfacial DMI) skyrmion. The nucleation of Néel skyrmions can occur in perpendicular materials also at zero field, while the nucleation of Bloch skyrmions needs an external out-of-plane field to lead the magnetization out of the sample plane, given that the materials where bulk DMI has been observed so far have an in-plane easy axis for the equilibrium magnetization at zero field.

The comparison of velocity-current are performed in devices realized by different materials depending on the skyrmion type. The different velocity-current slopes in Fig. 2a depend on the different physical parameters (as discussed above) and force related to the driving. This last aspect can be clearly seen by comparing the analytical expression describing velocity current relationship in [16] and the equation (2) of this paper (see also Supplementary Material S1).

Although the velocity-current curve of Bloch skyrmion motion driven by the SHE has the largest slope, this configuration is stable currents up to 15 MA/cm$^2$, at larger current the spin-orbit torque from the SHE gives rise either to the nucleation of domains[43] or magnetization reversal[7]. Therefore, we have concluded that, in the considered cases, the optimal strategy to design a skyrmion racetrack memory is scenario B. The physical limit which fix the maximal applicable current is given by the breathing mode of the Néel skyrmions (it can disappear) achieved in a transient time after the current for shifting the skyrmion is applied.

Furthermore, we have studied in detail the effect of the confining potential on the Néel skyrmion motion driven by the SHE at zero field we namely scenario B* stressing the fact that our results do not contradict the one presented in [21]. An additional study on the effect of the confinement in the case of the Bloch skyrmion driven by the STT has been presented in [44], in that case the STT originates from a current flowing transverse to the edge with the result that the skyrmion velocity is significantly enhanced by the confining potential (Fig. 2a in[44]) but it is stable only for a reduced current region. We have also performed simulations for a current transverse to the edge in the case of Bloch skyrmion motion driven by the SHE (scenario D*), finding out results (not shown) qualitatively similar to the ones achieved in scenario B*.

A comparison of the scenario B* with the scenario B (Fig. 3b and 4b) demonstrates that the threshold current is smaller in scenario B*, and that therefore the geometry and setup of B* can be used for ultra-low power storage devices. However, the skyrmion motion in scenario B is much more robust to thermal fluctuations and is much less sensitive to defects and edge roughness that are present in real devices. Skyrmion motion in scenario B can also be achieved for a wider current range and it is stable at larger currents implying larger skyrmion velocities.

In summary, we have analyzed the possible technological scenarios for controlling the shifting Néel or Bloch skyrmions driven by SHE or STT. Our results indicate that scenario B (Néel skyrmion motion driven by the SHE) is one of the most promising from a technological point of view and its experimental implementation could be achieved bearing in mind that the skyrmion mainly moves perpendicularly to the electrical current flow. Our results suggest a new route to design and develop a more efficient skyrmion racetrack memory.

## Acknowledgements

This work was supported by the project PRIN2010ECA8P3 from Italian MIUR, Project MAT2011-28532-C03-01 from Spanish government, and Project SA163A12 from Junta de Castilla y Leon. The authors thank Domenico Romolo for the graphical support and Praveen Gowtham for the careful reading of the manuscript. E.M. and L.T. acknowledge funding from the European Community under the 7th Framework Programme - The people Programme, Multi-ITN "WALL" Grant agreement no.: 608031.

## Author contributions

R. T. performed micromagnetic simulations. E. M. analyzed the data and supported the simulations. R. Z. developed the analytical theory. L. T. analyzed the data. G. F., R. T. and M. C. conceived and designed the numerical experiment, analyzed the data and wrote the paper. All authors contributed to the general discussion and comment on the manuscript.

## Additional information

Supplementary information is available in the online version of the paper. Reprints and permissions information is available online at www.nature.com/reprints. Correspondence and requests for materials should be addressed to G.F. (gfinocchio@unime.it).

## Competing financial interests

The authors declare no competing financial interests.

# Methods

## Micromagnetic model

The micromagnetic model is based on the numerical solution of the LLG equation[45] containing the STT[27] and the torque from SHE[7,11] ("state of the art" parallel micromagnetic solver GPMagnet[45,46,47,48]):

$$\frac{d\mathbf{m}}{d\tau} = -\mathbf{m} \times \mathbf{h}_{EFF} + \alpha_G \mathbf{m} \times \frac{d\mathbf{m}}{d\tau} + \frac{\mu_B P}{\gamma_0 e M_S^2}(\mathbf{j}_{FE} \cdot \nabla)\mathbf{m} - \frac{\mu_B P}{\gamma_0 e M_S^2}\beta \mathbf{m} \times (\mathbf{j}_{FE} \cdot \nabla)\mathbf{m}$$
$$- \frac{g\mu_B}{2\gamma_0 e M_S^2 L}\theta_{SH} \mathbf{m} \times \mathbf{m} \times (\hat{z} \times \mathbf{j}_{HM}) \quad (3)$$

where $\mathbf{m}$ and $\mathbf{h}_{EFF}$ are the normalized magnetization and the effective field of the ferromagnet. $\tau$ is the dimensionless time. The effective field includes the standard magnetic field contributions and the DMI field [22]. $\alpha_G$ is the Gilbert damping, $\mu_B$ is the Bohr Magneton, $P$ is the polarization coefficient of the in-plane electrical current, $\gamma_0$ is the gyromagnetic ratio, $e$ is the electron charge, $M_s$ is the saturation magnetization, $\mathbf{j}_{FE}$ is the in-plane electrical current flowing through the ferromagnet, $\beta$ is the spin-torque non-adiabatic factor, $g$ is the Landè factor, $\theta_{SH}$ is the spin-Hall angle, $L$ is the thickness of the ferromagnetic layer, $\hat{z}$ is the unit vector of the out-of-plane direction and $\mathbf{j}_{HM}$ is the in-plane current injected via the heavy metal. We use a discretization cell of 2x2x1nm$^3$ for the scenarios A and B (CoFeB alloy) and 0.5x0.5x1nm$^3$ for the scenarios C and D (MnSi), which are smaller than the respective exchange length $l_{ex}$ and DMI length $l_D$ (see Table 1 for all the physical and geometrical parameters). Although for the scenario B we have described in details the results for the parameters reported in Table 1, we have also achieved qualitative similar results at zero external field for thicknesses 1.0nm ($\alpha_G$=0.03, $M_S$=1000 kA/m, $K_u$=0.80 MJ/m$^3$, $A$=20.0 pJ/m, $D$=1.8 mJ/m$^2$), 0.8 nm ($\alpha_G$=0.05, $M_S$=900 kA/m, $K_u$=0.85 MJ/m$^3$, $A$=20.0 pJ/m, $D$=2.7 mJ/m$^2$), and 0.6 nm ($\alpha_G$=0.1, $M_S$=800 kA/m, $K_u$=0.95 MJ/m$^3$, $A$=15.0 pJ/m, $D$=3.3 mJ/m$^2$).[49,50,51,52,53] The value of $D$ gives rise to a skyrmion diameter of 46 nm. We have also analyzed the same scenario studied in [21] (the heavy metal is Pt, the ferromagnet is Co with a thickness of 0.4 nm, $\alpha_G$=0.3, $M_S$=580 kA/m, $K_u$=0.8 MJ/m$^3$, $A$=15.0 pJ/m, $D$=3.0 mJ/m$^2$), obtaining the same qualitative results described here, but now the skyrmion diameter is around 24 nm.

The bulk DMI energetic density expression is[22]:

$$\varepsilon_{BulkDMI} = D\mathbf{m} \cdot \nabla \times \mathbf{m} \quad (4)$$

and, making the functional derivative of equation (4), the bulk DMI effective field is derived:

$$\mathbf{h}_{BulkDMI} = -\frac{1}{\mu_0 M_S}\frac{\delta \varepsilon_{BulkDMI}}{\partial \mathbf{m}} = -\frac{2D}{\mu_0 M_s}\nabla \times \mathbf{m} \qquad (5)$$

being *D* the parameter taking into account the intensity of the DMI. The boundary conditions related to the bulk DMI are expressed by[22]:

$$\frac{d\mathbf{m}}{dn} = \frac{\mathbf{m}\times\mathbf{n}}{\xi} \qquad (6)$$

where **n** is the unit vector normal to the edge and $\xi = \frac{2A}{D}$ (being *A* the exchange constant) is a characteristic length in presence of DMI.

The interfacial DMI energetic density, effective field expression and boundary conditions are[22]:

$$\varepsilon_{InterDMI} = D\left[m_z \nabla\cdot\mathbf{m} - (\mathbf{m}\cdot\nabla)m_z\right] \qquad (7)$$

$$\mathbf{h}_{InterDMI} = -\frac{1}{\mu_0 M_S}\frac{\delta \varepsilon_{InterDMI}}{\partial \mathbf{m}} = -\frac{2D}{\mu_0 M_S}\left[(\nabla\cdot\mathbf{m})\hat{z} - \nabla m_z\right] \qquad (8)$$

$$\frac{d\mathbf{m}}{dn} = \frac{1}{\xi}(\hat{z}\times\mathbf{n})\times\mathbf{m} \qquad (9)$$

where $m_z$ is the *z*-component of the normalized magnetization and the ultra-thin film hypothesis $\left(\frac{\partial \mathbf{m}}{\partial z} = 0\right)$ is considered.

The thermal effect is included as an additional stochastic term of the effective field computed as:

$$\mathbf{h}_{th} = (\xi/M_s)\sqrt{2(\alpha_G K_B T / \mu_0 \gamma_0 \Delta V M_s \Delta t)} \qquad (10)$$

where $K_B$ is the Boltzmann constant, $\Delta V$ is the volume of the computational cubic cell, $\Delta t$ is the simulation time step, *T* is the temperature and ξ is a Gaussian stochastic process. The thermal field $\mathbf{h}_{th}$ satisfies the following statistical properties:

$$\begin{cases}\langle h_{th,k}(\vec{r},t)\rangle = 0 \\ \langle h_{th,k}(\vec{r},t) h_{th,l}(\vec{r}\,',t')\rangle = F\delta_{kl}\delta(t-t')\delta(\vec{r}-\vec{r}\,')\end{cases} \qquad (11)$$

where *k* and *l* represent the Cartesian coordinates *x*, *y*, *z*. According to that, each component of $\mathbf{h}_{th} = (h_{th,x}, h_{th,y}, h_{th,z})$ is a space and time independent random Gaussian distributed number (Wiener process) with zero mean value[54, 55]. The constant *F* measures the strength of thermal fluctuations and its value is obtained from the fluctuation dissipation theorem.

| Parameters | Scenario A | Scenario B | Scenario C | Scenario D |
|---|---|---|---|---|
| Skyrmion type | Néel | Néel | Bloch | Bloch |
| Motion Source | STT | SHE | STT | SHE |
| Ferromagnetic Material | CoFeB | CoFeB | MnSi | MnSi |
| Heavy Metal[38, 40, 41] | W | W | - | W |
| Saturation magnetization $M_s$ ($10^3$ A/m)[49] | 1000 | 1000 | 152 | 152 |
| Exchange constant $A$ ($10^{-12}$ J/m)[16, 17] | 20 | 20 | 0.32 | 0.32 |
| Perpendicular anisotropy constant $K_u$ ($10^6$ J/m$^3$)[38, 49] | 0.8[§] | 0.8[§] | - | - |
| Gilbert damping $\alpha_G$ [17,49] | 0.015 | 0.015 | 0.04 | 0.04 |
| Non-adiabatic parameter $\beta$ | 0.015 | - | 0.04 | - |
| DMI parameter $D$ ($10^{-3}$ J/m$^2$)[16, 17, 21, 41, 50] | 4 | 4 / 1.8[#] | 0.115 | 0.115 |
| Polarization coefficient $P$[51] | 0.4 | 0 | 0.4 | 0 |
| Spin-Hall angle $\theta_{SH}$ [38, 52] | 0 | -0.33 | 0 | -0.33 |
| Magnetic field $H_{ext}$ (mT) | 250 | 250 / 0[#] | 250 | 250 |
| Temperature $T$ (K) | Room | Room | <30 | <30 |
| Exchange length $l_{ex}$ (nm) | 5.6 | 5.6 | 4.7 | 4.7 |
| DMI length $l_D$ (nm) | 10 | 10 | 5.56 | 5.56 |

**Table 1 | Summary of the physical parameters used in the micromagnetic simulations.** [#]In the case of Scenario B, we have used a different value of DMI parameter for the two different external fields to maintain the skyrmion size constant. [§]The perpendicular anisotropy is achieved via the electrostatic interaction between the Fe atoms of the CoFeB with the O atoms of the oxide MgO on top. The relation $\alpha_G = \beta$ has been used as a good compromise among the possible $\beta$ values.

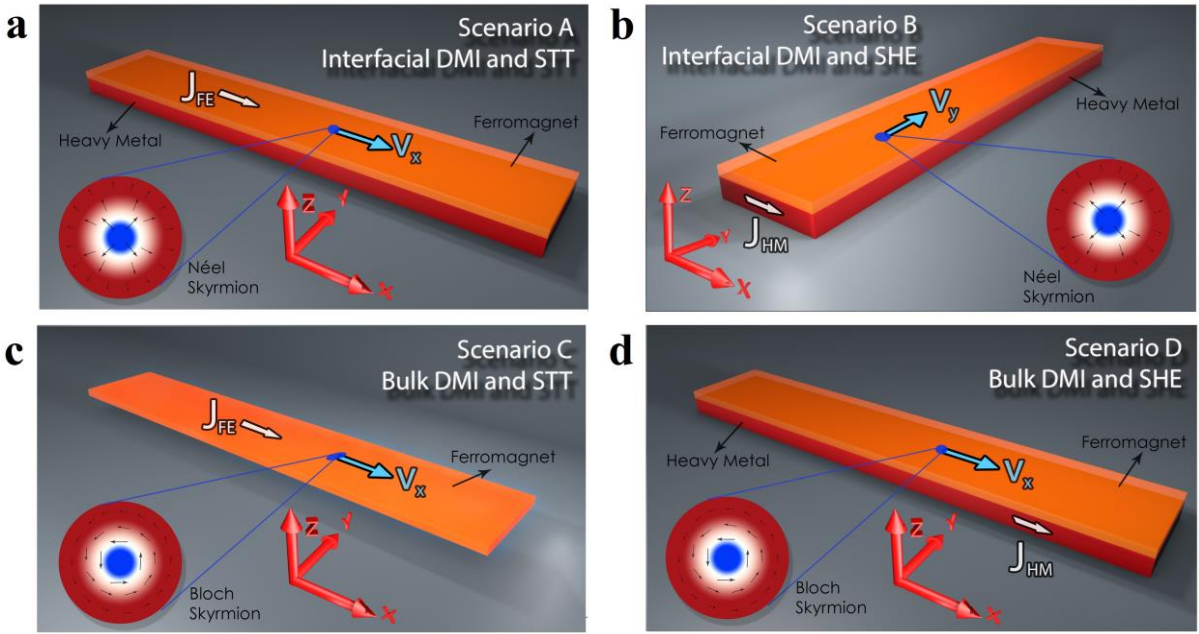

**Figure 1 | Four different scenarios for the design of a skyrmion racetrack memory. a**, Néel skyrmion motion driven by the STT. **b**, Néel skyrmion motion driven by the SHE. **c**, Bloch skyrmion motion driven by the STT. **d**, Bloch skyrmion motion driven by the SHE. The four insets show the spatial distribution of the Néel and Bloch skyrmion, where the background colors refer to the *z*-component of the magnetization (blue negative, red positive), while the arrows are related to the in-plane components of the magnetization. The current flows along the *x*-direction. The skyrmion moves along the *x*-direction in the scenarios A, C, and D and along the *y*-direction in the scenario B.

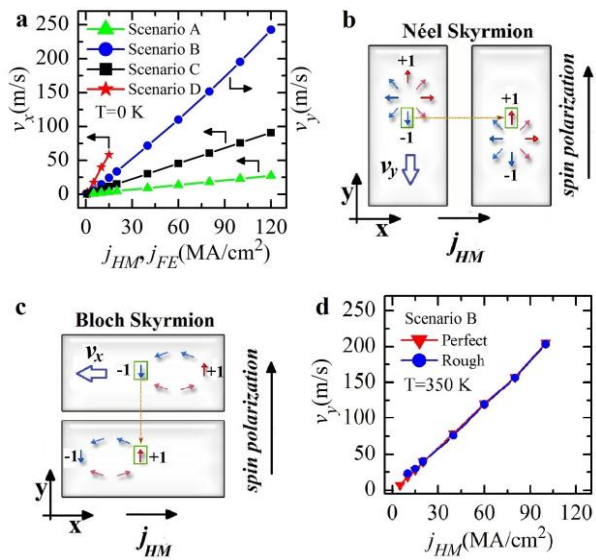

**Figure 2 | Velocity-current curves for the skyrmion motions. a**, A comparison among the skyrmion velocities obtained for each scenario (A, B, C, and D). The current $j_{FE}$ is related to the scenarios A and C while $j_{HM}$ to the scenarios B and D. **b**, Sketch of the motion mechanism of the Néel skyrmion driven by the SHE along the *y*-direction. **c**, Sketch of the motion mechanism of the Bloch skyrmion driven by the SHE along the *x*-direction. **d**, Skyrmion velocities as a function of $j_{HM}$ obtained for the scenario B: (i) thermal fluctuations (*T*=350 K) and perfect strip (red curve) and (ii) thermal fluctuations (*T*=350 K) and rough strip (blue curve). The arrows for **b** and **c** refer to the in-plane components of the magnetization, the spin-polarization of $j_{HM}$ is also displayed.

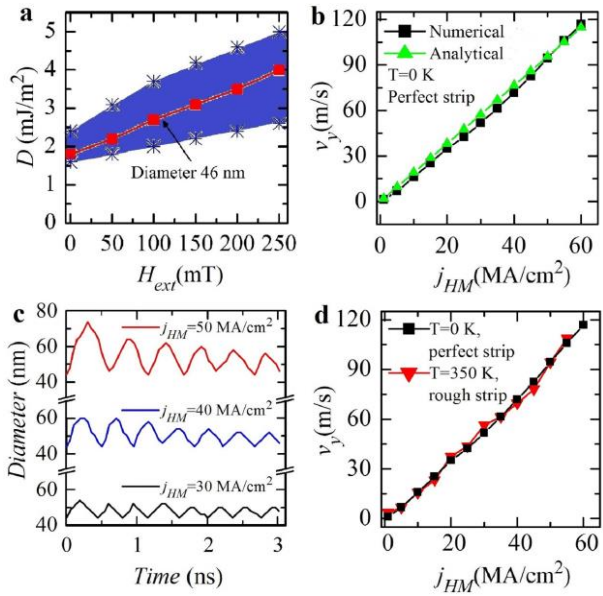

**Figure 3 | Skyrmion racetrack memory at zero bias field. a**, Phase diagram (Interfacial DMI parameter vs. external field) of the skyrmion stability. The colored part highlights the region where the skyrmion is stable and the red curve points out the values of $D$ and $H_{ext}$ for which the skyrmion diameter is 46 nm. **b**, Skyrmion velocity as a function of the current calculated by micromagnetic (black curve) and analytical computations (green line) (the parameters can be found in Supplementary Material S1). **c**, Time domain evolution of the skyrmion diameter during the transient breathing mode at three different values of the $j_{HM}$, 30 MA/cm$^2$ (black curve), 40 MA/cm$^2$ (blue curve) and 50 A/cm$^2$ (red curve). **d**, A comparison of skyrmion velocity as a function of the current for an ideal strip ($T$=0K) and a rough strip and room temperature.

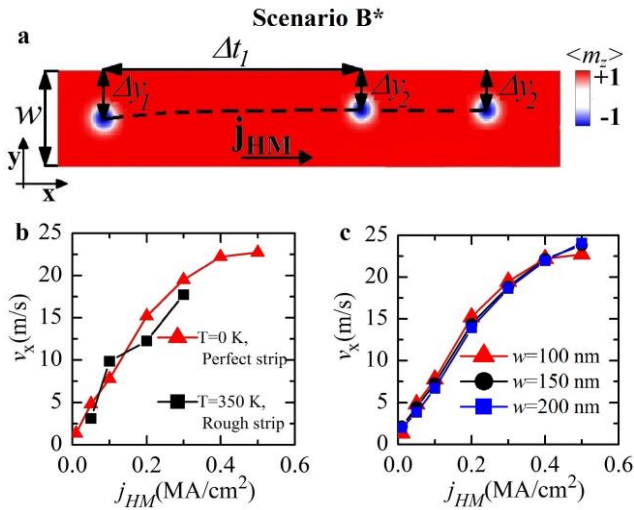

**Figure 4 | Néel skyrmion motion due to the SHE: effect of the confinement. a**, Sketch of the skyrmion motion due to the SHE in a sample confined along the $y$-direction. **b**, Skyrmion velocity ($x$-direction) as a function of the current: (i) no thermal fluctuations and perfect strip (red curve) and (ii) thermal fluctuations ($T$=350 K) and rough strip (black curve). **c**, Skyrmion velocity as a function of the current when the strip width $w$ is 100 nm (red curve), 150 nm (black curve) and 200 nm (blue curve). The colors for **a** refer to the $z$-component of the magnetization (blue negative, red positive).